# Probing into gas leakage characteristics of ventilated supercavity through bubbly wake measurement

Siyao Shao[1,2], Jiaqi Li[1,2], Kyungduck Yoon[3], and Jiarong Hong[1,2,*]

[1] St. Anthony Falls Laboratory, University of Minnesota, Minneapolis, MN 55414, USA
[2] Department of Mechanical Engineering, University of Minnesota, Minneapolis, MN 55455, USA
[3] Department of Mechanical Engineering, Georgia Institute of Technology, Atlanta, GA 30332, USA,

**Abstract**

The stability of ventilated supercavitation is strongly influenced by gas leakage characteristics of the cavity. In this study we conduct a systematic investigation of such characteristics under different closure conditions including re-entrant jet (RJ), quad vortex (QV), twin vortex (TV), and pulsating twin vortex (PTV), generated from different tunnel speeds and ventilation conditions. Using high speed digital inline holography (DIH), all the individual bubbles shed from the cavity are imaged downstream and are used to quantify the instantaneous gas leakage from the cavity. In general, the supercavity gas leakage exhibits significant fluctuations under all closure types with the instantaneous leakage rate spiking up to 20 times of the ventilation input under RJ and QV closures. However, the magnitude and occurrence rate (frequency) of such excessive gas leakage (above ventilation input) vary substantially across different closures, tunnel speeds and ventilation conditions. Particularly, as the supercavity transitions from RJ, to QV, TV, and PTV with increasing ventilation or decreasing tunnel speed, the relative excessive gas leakage decreases sharply from RJ to QV, plateaus from QV to TV, and drops again from TV to PTV. Correspondingly, the occurrence frequency of such excessive leakage first exhibits a double peak distribution under RJ, migrates to a single peak mode under QV and TV, and eventually transitions to a distribution with a broadened peak at a higher frequency under PTV. These trends can be explained by the flow instabilities associated with three gas leakage mechanisms, i.e., re-entrant jet impingement leakage, vortex tube gas leakage, and cavity pulsation induced bubble pocket shed-off, whose relative significance changes under different closure and flow conditions. Subsequently, two metrics are introduced to quantify the relative change of ventilation needed to compensate the change of extra gas loss and the predictability of the occurrence of excessive gas leakage, respectively. Based on these metrics, we suggest that the supercavity operating under TV closure with moderate ventilation is optimal for ventilation-based controls of supercavity stability.

## 1. Introduction

Ventilated supercavitation, i.e., a special case of cavitation which the cavitating body can be enclosed in a gas bubble generated by injecting gas behind a cavitator, has gained substantial attention for its potential of drag reduction in high-speed underwater applications (Campbell and Hirborne 1958, Epshtein 1975, Franc and Michel 2006, Kinzel et al. 2009). One of most critical problems to implement such technique is to stabilize the ventilated supercavity under the various unsteady conditions experienced by a supercavitating device during its practical operation. Specifically, the ventilated supercavity exhibits substantial variation of dimension and topology as observed by Lee et al. (2013), Karn et al. (2015a), and Shao et al. (2018). Such behaviors of supercavity can result in a supercavitating object penetrating the cavity surface and adversely affecting the drag reduction capability of the supercavitation (Sanabria et al. 2015). Conventional fin-based control causes additional planing force limiting the operating speed of the supercavitating vehicles (Kirshner et al. 2002). An alternative controlling strategy is to adjust ventilation conditions such as ventilation gas content and ventilation ports arrangement to compensate unsteady effects on cavity morphology or make up excessive gas leakage (e.g., using ventilation to suppress supercavity pulsation as proposed by Skidmore 2016). Such controlling strategy is based on the recent observations of the influence of the ventilation on the supercavity behaviors (Kawakami & Arndt 2011, Karn et al. 2016a, Karn et al. 2016b, Kinzel et al. 2017, Sun et al. 2019, and Sun et al. 2020). Particularly, Kawakami & Arndt (2011) has observed that the cavity dimension varies substantially with the change of the amount of gas injected into the cavity. Karn et al. (2016a) systematically investigated the supercavity topology (i.e., cavity closure) under different experimental conditions and found the supercavity closure pattern variations with the change of gas ventilation and flow speed. Specifically, the vortex-based closure cavity is formed by small pressure difference across the cavity closure region. With increasing pressure difference across the closure region, the supercavity transitions to re-entrant jet closure type. In addition, vortex-based closure supercavity formed with high ventilation rate can be sustained under lower ventilation with a re-entrant jet closure. The theoretical analysis by Karn et al. (2016a) provides a framework connecting the supercavity closure change with the change of gas ventilation and flow speed. The follow-up study (Karn et al. 2016b) showed that the frequency and amplitude of unsteady flow alter the gas leakage and the correspond ventilation required to form and sustain a supercavity non-monotonously. Subsequently, Sun et al. (2019) found that increasing ventilation to a supercavity according to the ambient unsteady flow conditions (i.e., ventilates more air into the cavity under higher amplitude flow unsteadiness) reduces the load on the cavitating body and compensates the cavity shape change due to the extra gas leakage associated with the unsteady flow. As pointed out by Kinzel et al. (2017) using numerical simulation, the change of cavity behaviors can be attributed to the interaction between the gas ventilation and ambient water flow near the gas-liquid interface. Therefore, a supercavity encountering stable flow conditions can also exhibit unsteady gas leakage due to the instabilities developed near gas-liquid interface. As an example, the gas pocket shed-off from the surface of ventilated supercavity has been observed by Sun et al. (2020) under high ventilation rate with stable flow conditions. These observations indicate that the ventilation-based supercavity control should consider the time-varying (not just the average) gas leakage from the supercavity. However, no such investigation has been conducted to understand the temporal variation in gas leakage of a supercavity under various flow conditions.

For over a half century, most of the prior investigations of the supercavity gas leakage have been conducted in an average sense to establish semi-empirical correlations of the gas leakage rate with flow conditions described by dimensionless parameters and cavity geometries. Specifically, the ventilation input and gas leakage rate are usually quantified by the non-dimensional volumetric gas rate coefficient, $C_Q = Q/U_\infty d_C^2$, where $Q$ is the volumetric ventilation rate (Note that in the current study, we use the volumetric air flow rate under standard condition, i.e., $Q_{AS}$, to define gas rate coefficient $C_{Qs}$), $U_\infty$ is the tunnel speed, and $d_C$ refers to the cavitator diameter. The flow conditions are usually quantified by the cavitation number, $\sigma_C = 2(P_\infty - P_C)/(\rho_W U_\infty^2)$, and Froude number, $Fr = U/\sqrt{gd_C}$, where $P_\infty$ and $P_C$ are the upstream test-section pressure and the supercavity pressure, respectively, $\rho_W$ is the water density, $g$ denotes the gravitational acceleration. These early works characterize two gas leakage mechanisms, i.e., gas leakage via vortex tubes formed at the rear of the supercavity due to gravity under low $Fr$ (Cox & Clyden 1955, Campbell & Hilborne 1958, Logvinovich 1969, and Epshtein 1973), and air leakage via internal boundary layer formed at the gas-liquid interface under high $Fr$ (Spurk 2002). To further investigate the underlying gas leakage physics, Kinzel et al. (2009) and (2017) combined the internal boundary layer gas leakage theory of Spurk (2002) and vortex-tube gas leakage using numerical simulation. The study pointed out the connection between the internal flow in the center portion of the cavity, the shear-layer gas leakage and eventually the amount of gas leaked from the vortex tube at the closure region. Through a framework established upon non-uniform cavity internal pressure distribution associated with internal flow, Karn et al. (2016a) pointed out the influence of cavity closure types on the gas leakage. Specifically, the large pressure difference across the cavity surface near closure region causes the re-entrance of water jet into the cavity to form a re-entrant jet (RJ) cavity closure. Under RJ closure, a toroidal vortex comprising of gas-liquid mixture formed near closure region facilitate the leakage of ventilation gas. For the vortex-based closure formed with low pressure difference across the cavity surface, the gas leakage is through the vortex tube formed at the cavity closure. Recently, Wu et al. (2019) elucidated the connection between the internal boundary layer air entrainment and the internal flow in the center portion of the supercavity experimentally through internal flow visualization and measurement. Additionally, Wu et al. observed the coexistence of the vortex tube gas leakage and toroidal vortex gas leakage due to the liquid entrainment into the cavity near closure region on a RJ closure cavity. With increasing of ventilation from RJ cavity to TV cavity, the toroidal gas leakage gradually diminishes with elongation of the cavity and the widening of the vortex tubes behind the cavity. At the same time, the supercavity elongates which allows more gas leakage through the internal boundary layer and eventually the vanishing of gas leakage through toroidal vortex. From the observation of partial cavity formed by a back step cavitator, Qin et al. (2019) and Yoon et al. (2020) found that even with steady tunnel speed and air ventilation, the cavity exhibits fluctuations of gas leakages without quantifications.

Consequently, a systematic investigation of temporal characteristics of gas leakage is necessary to establish a comprehensive framework for ventilation-based control. However, such investigation encounters immense technical difficulties. Measuring internal flow to calculate the instantaneous gas leakage is limited to very few conditions and the limited windows of such measurements could not cover whole internal flow field especially the internal boundary layers which could cause large uncertainties (Wu et al. 2019). Therefore, we propose an alternative approach to measure the velocity and size distribution of the bubbles in the wake of the cavity to infer the gas leakage rate. The similar measurements have been demonstrated previously by calculating the average volumetric flow rate of air aerated through an aeration hydrofoil using shadowgraphic imaging (Karn et al. 2015b, Karn et al. 2015c). The 3D imaging technique are preferred over the shadowgraph since the latter has limited depth-of-field and the out-of-focus bubbles will cause large uncertainties in the calculation of instantaneous gas leakage. In the recent decade, digital inline holography (DIH) has emerged as a low-cost and compact tool for 3D measurements of particle size distributions and velocities (Katz & Sheng 2010). DIH employs a coherent light source (e.g., laser) and a single camera to record the interference pattern (i.e., hologram) generated from the scattered light of an object and the non-scattered portion of the illumination light source (Katz & Sheng 2010). The holograms are reconstructed digitally, and the particle size and locations are extracted through a segmentation process from the reconstructed optical field. The particle velocities are subsequently measured using particle tracking. Compared to conventional imaging approaches, DIH could provide 3D high resolution particle measurements without assumptions. Such technique has been demonstrated in the measurement of bubble field with extended depth-of-field to get accurate quantification of gas leakage rate from a supercavity (Shao et al. 2019). In the current study, we apply DIH techniques to measure the bubble size distribution in the supercavity wake over a broad range of flow conditions to infer the temporal change of supercavity gas leakage and investigate the flow physics governing dynamic features of supercavity gas leakage.

The paper is structured as the following: Section 2 provides a description of the experimental methods and data process procedures. Section 3.1 reports the results of gas leakage characteristics across different cavity closures and the physical interpretation of our findings. Section 3.2 summarizes the gas leakage characteristics across different tunnel speeds. Subsequently, section 3.3 compares the effects of ventilation rates on gas leakage characteristics. Section 3.4 connects the experimental results with ventilation-based control. Section 4 is a summary and discussion of the results.

## 2. Experimental Method

The experiments are conducted at the high-speed water tunnel in the Saint Anthony Falls Laboratory (SAFL). The water tunnel has a test section of 1200 mm (length) × 190 mm (width) × 190 mm (height) and is capable of operating at flow speed up to 20 m/s with a turbulence level of 0.3%. A large dome-shaped settling chamber is situated at the upstream of the test section with the capability of fast removal of gas bubbles allowing a continuous operation of cavitation and ventilation experiments. In the recent years, this facility has been used for a number of supercavitation experiments (Karn et al. 2016a and b, Shao et al. 2017, Shao et al. 2018). A backward facing cavitator (shown in the inset figure of Fig. 1) of 10 mm-in-diameter is employed to generate a free standing ventilated supercavity for the experiments (detailed information provided in Karn et al. 2016a). The small cavitator used in the current study minimizes the ventilation requirement for varying closure types. The cavitator is mounted at the trailing edge of a hydrofoil strut of 5 mm maximum thickness to minimize the influence from the wake of the mounting strut. As shown in Fig.1, the inline holographic system is placed at 405 mm downstream of the cavitator. The choice of DIH measurement location considers both the dimension of water tunnel test section and the expansion of bubbly wake to obtain more accurate estimate of gas leakage through the measurements of bubble volume. On the one hand, the measurement location is set to be sufficiently downstream to ensure individual bubbles in the bubbly wake to be fully dispersed to avoid occlusion of bubbles which lowers the accuracy of bubble volume estimate using our DIH imaging system. On the other hand, the measurement location needs to be close to the supercavity to avoid the expansion of bubbly wake beyond the measurement volume causing underestimate of total bubble volume and gas leakage rate. Additionally, the aluminum supporting structures downstream of the water tunnel can potentially block the optical path of our imaging system if the measurement window is placed further downstream. After trials and errors, we determined that 405 mm downstream of the cavitator is the optimal location for measurement window in our experiments. The holograms are recorded with a high-speed DIH setup comprised of a 532 nm continuous diode laser, beam expansion optics, and a NAC HX-5 high speed camera with a Nikon 105 mm imaging lens. The pixel resolution of the hologram sequence is 54.8 μm/pixel. The high-speed camera is operated from 3000 frames per second (fps) to 10,000 fps depending on the tunnel speed and measurement tasks. The image size is set to be 512 pixel × 512 pixel (i.e., a field-of-view of 28 mm × 28 mm). The sample hologram is shown here as Fig. 2. During the course of experiments, the air flow rate is regulated by an FMA-2609A mass flow controller with a unit of standard liter per minute (SLPM), which is the volumetric flow rate at the standard temperature (273.15 K or 0 ºC) and standard pressure (101.3 kPa or 1 atm). The uncertainty in the air flow rate measurement is ±1 % with a full-scale reading up to 40 SLPM. As Fig. 1 shows, both test section pressure and cavity internal pressure ports are situated at 1.4 m above the ground. The test section pressure is monitored by a Rosemount 3051S pressure sensor and the differential pressure across the cavity surface is measured by a Validyne DP-15 pressure sensor with one side connected with the test section pressure port and another side connected with the cavity pressure port. Additionally, a pressure port is located at the settling chamber where the water flow has zero speed. Along with the test section pressure port, the differential pressure across the settling chamber and test section is measured using a Rosemount 3051S sensor to calculate flow speed during the experiments. The standard errors of the pressure measurements are around 0.1 kPa for both Rosemount 3051S pressure sensors which yields a maximum error of 0.11 m/s in the result of instantaneous flow speed and a mean error around 0.02 m/s. The Validyne DP-15 pressure transducer has a standard error of 0.1 kPa for the $\sigma_C$ measurements. Overall, in the present experiments, the maximum systematic uncertainties of the $C_{Qs}$, $Fr$ and $\sigma_C$ are around 2%. A YSI thermistor with ±0.2 °C uncertainty is used to measure the local water temperature. During the experiments, the temperature was measured to be around 20 °C.

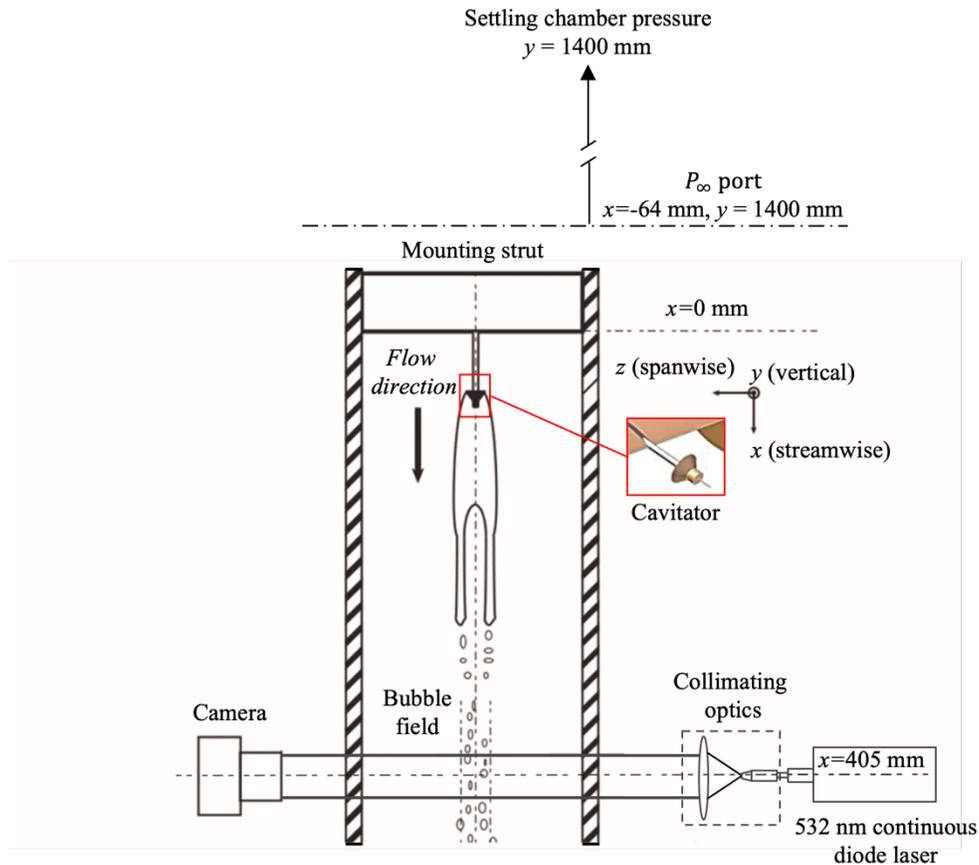

Figure 1. Schematics showing the generation of ventilation supercavity and the corresponding digital inline holography (DIH) setup for supercavity wake bubble measurements in the test section.

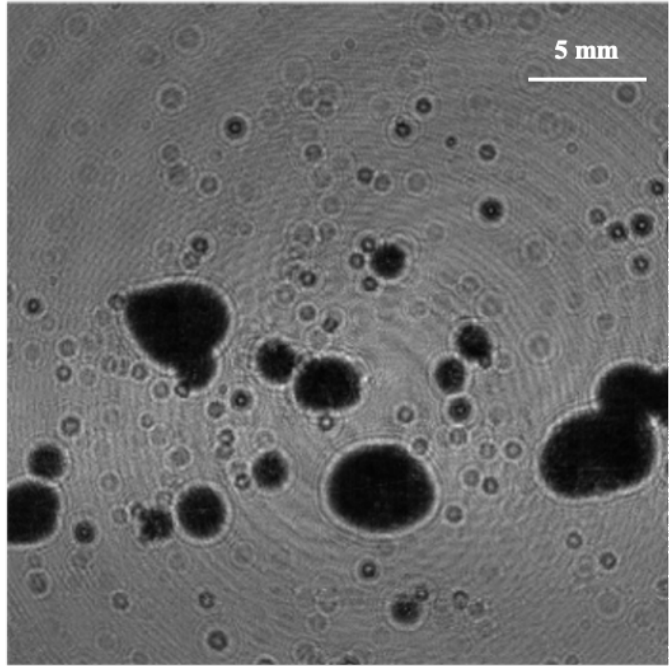

Figure 2. Sample bubble hologram captured in the current study.

The hologram processing has following steps. First, we employ the conventional hologram processing approach proposed by Shao et al. (2019) to process 1000 holograms to establish a database of bubble size, shape, and enhanced holograms and corresponding labels. Subsequently, we use such database as a training dataset to train a machine learning model for batch processing holograms following the learning-based hologram processing method proposed by Shao et al. (2020). The resulted bubble location information is used to conduct a particle tracking to resolve the velocities of individual bubbles (Crocker & Grier 1996). The estimation of bubble volume from its 2D projections obtained from DIH measurements follows the procedures proposed by Karn et al. (2015b). The sizes of bubbles are quantified by the semi-major ($a$) and semi-minor axis length ($b$) of elliptical fitting results of each bubble. The ellipse fitting of each bubble is rotated around its major axis to obtain a prolate spheroid or an ellipsoid of revolution. The volume of individual bubbles can be mathematically determined as $(4\pi/3)b^2 a$. It is worth noting that such an approach to estimate bubble volume yields maximum uncertainty around 5% for highly irregular bubbles such as dumbbell shapes through theoretical analysis. Additionally, with relatively high tunnel speed (>2 m/s), the bubbles are well-aligned with the direction of tunnel speed and the uncertainty caused by the bubble inclination with respect to the recording plane is much smaller than the temporal fluctuation of the gas leakage (Shao et al. 2019). The last step is to calculate the gas leakage rate following Eqn. 1 and Eqn. 2 (Karn et al. 2015b).

$$Q_{\text{GL}} = \sum_{i=1}^{N} \left(\frac{u_i}{w}\right)\left(v_i \times \frac{P_\infty}{P_0} \times \frac{T_0}{T_\infty}\right) \quad (1)$$

$$v_i = (4\pi/3)a_i b_i^2 \quad (2)$$

where $N$ denotes number total number of bubbles in a single image, $w$ refers to the width of a single bubble image (28 mm). $u_i$ and $v_i$ signify velocity and volume of individual bubbles, respectively. Finally, $P_0$ and $T_0$ correspond to standard temperature and pressure, and $P_\infty$ and $P_\infty$ are corresponding test-section conditions. The infinity sign in the subscripts of pressure and temperature represents of the free stream test section variables (i.e., $P_\infty$ and $T_\infty$) with their measurement locations situating sufficiently far away from the supercavity in the test section (i.e., downstream of the cavity closure near the DIH measurement window in this case). Despite a strong fluctuation of instantaneous gas leakage rate ($Q_{GL}$) shown in Fig. 3(b), the averaged gas leakage from our measurement agrees well with the input ventilation for all the experimental data points (note that some data points are overlapped due to closely matched gas leakage estimation from bubble data) (Fig. 3c). Note that with increasing ventilation and the transition of the cavity closure from re-entrant jet to vortex-based, the chance of individual large bubbles occluding other bubbles increases, resulting in an 10% maximum underestimate of gas leakage using total bubble volumes derived from our holographic imaging. Overall, the uncertainty caused by bubble overlaps and shape deviation from ellipsoidal shape is limited to 12% based on our estimation, significantly lower than the observed gas leakage fluctuations (see the detailed analysis in the Section 3 and 4). Last, in order to compare the results across different cases, the instantaneous gas leakage rate is further normalized by the following Eqn. 3.

$$\hat{Q} = \frac{Q_{GL}}{Q_{AS}} \quad (3)$$

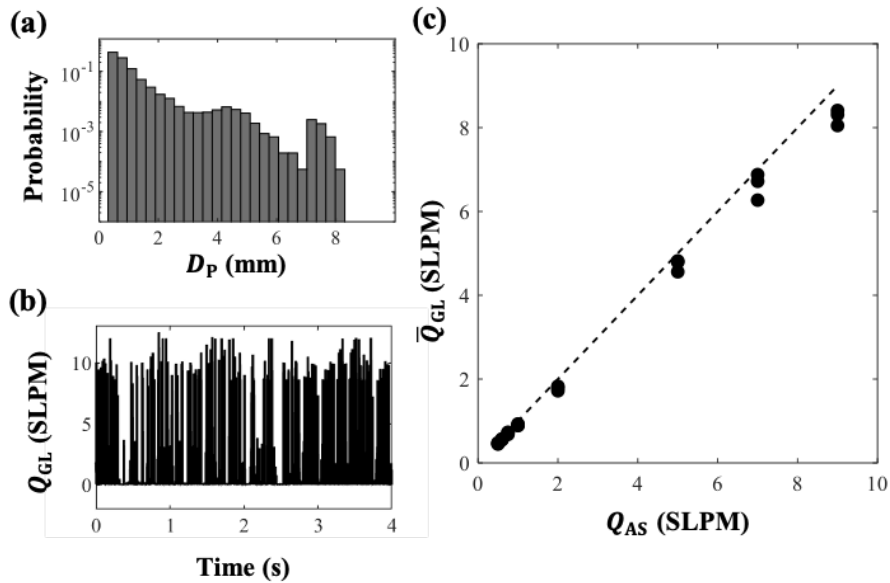

Figure 3. (a) Sample histogram of the extracted bubble size distribution from DIH measurements (corresponding to 4 s data under ventilation rate at 0.5 SLPM with tunnel speed at 3 m/s) and (b) the corresponding instantaneous gas leakage rate derived from the bubble measurements. Note that the particle size shown in (a) is quantified using area-equivalent diameter of the bubbles. (c) The comparison of ventilation input and average ventilation rate derived from DIH bubble measurements for all the experimental conditions.

According to Karn et al. (2016a), we set the tunnel speed from 3 m/s to 5 m/s and ventilation rate from 0.5 SLPM to 9 SLPM during the current experiments. Such conditions allow us to cover four major closure modes, i.e., re-entrant jet (RJ), quad-vortex (QV), twin-vortex (TV), and pulsating twin vortex (PTV), during the experiments. The high-speed imaging at a constant frame rate of 10,000 fps is also conducted near the closure region to determine the closure modes and temporal variations of supercavity closure under different experimental conditions. The experimental conditions and supercavity closure types are summarized in Tab. 1. Note that we did not observe sustenance of QV closure (existing for longer than four seconds) under $Fr = 16.0$ for continuous wake bubble measurements since the QV closure is susceptible to the fluctuations of differential pressure across the cavity closure region especially for high tunnel speed case with low test section pressure (Karn et al. 2016a). For each condition, three sets of DIH experiments are conducted independently with each set spanning a time duration of 8 seconds (i.e., over 24,000 independent frames) to ensure statistical convergence during the analysis. It should be noted that the present paper use probability distributions of different gas leakage statistics in the results section to reveal and analyze the gas leakage characteristics and its underlying mechanisms under different experimental conditions.

| Closure type | $U_\infty$ (m/s) | $Fr$ | $Q_{AS}$ (SLPM) | $C_{Qs}$ |
|---|---|---|---|---|
| Re-entrant jet (RJ) | 3.0 | 9.6 | 0.5 | 0.028 |
|  | 4.0 | 12.8 | 0.5, 0.6, 0.75 | 0.021, 0.025, 0.031 |
|  | 5.0 | 16.0 | 0.5, 0.6, 0.75, 1.0 | 0.017, 0.020, 0.025, 0.033 |
| Quad vortex (QV) | 3.0 | 9.6 | 0.6, 0.75 | 0.033, 0.042 |
|  | 4.0 | 12.8 | 1.0 | 0.042 |
| Twin vortex (TV) | 3.0 | 9.6 | 1.0, 2.0 | 0.056, 0.11 |
|  | 4.0 | 12.8 | 2.0, 5.0 | 0.083, 0.21 |
|  | 5.0 | 16.0 | 2.0, 5.0, 7.0 | 0.067, 0.17, 0.23 |
| Pulsating twin vortex (PTV) | 3.0 | 9.6 | 5.0, 7.0, 9.0 | 0.28, 0.39, 0.50 |
|  | 4.0 | 12.8 | 7.0, 9.0 | 0.29, 0.38 |
|  | 5.0 | 16.0 | 9.0 | 0.30 |

Table. 1. Summary of experimental conditions including closure type, tunnel speed ($U_\infty$) and ventilation rate ($Q_{AS}$) for DIH bubble measurements.

## 3. Results

### 3.1. Temporal characteristics of gas leakage across different closure types

The instantaneous gas leakage of supercavity is first compared qualitatively across different closure types under the same tunnel speed ($U_\infty = 3.0$ m/s) and Froude number, i.e., $Fr = 9.6$ (Fig. 4). As discussed in Karn et al. (2016a), the dimensionless numbers such as Froude number and gas rate coefficient cannot capture the detailed characteristics (e.g., closures) of the supercavity beyond the overall cavity dimensions. Therefore, we use both conventional dimensionless numbers in the figures and figure captions, which makes it easier for the readers to relate our experimental results with the ones from different studies. As shown in the figure, the instantaneous gas leakage rate exhibits large fluctuations (i.e., intermittent gas leakage spikes) for all cavity closures. Specifically, the gas leakage of RJ closure cavity frequently exceeds the ventilation input ($\hat{Q}>1$) and exhibits large spikes with magnitude reaching more than 20 times of the ventilation input on some occasions (Fig. 4a). As the ventilation increases, the cavity transitions to QV mode/closure and the gas leakage fluctuation reduces both in frequency and magnitude (Fig. 4b). With further increase of ventilation, the frequency and magnitude of gas leakage keeps decreasing as the cavity transitions to TV (Fig. 4c). Lastly, at the highest ventilation, the cavity transitions to PTV closure and the gas leakage fluctuates more frequently (demonstrated in the inset figure of Fig. 4d) but at a significantly lower amplitude in comparison to other closure conditions.

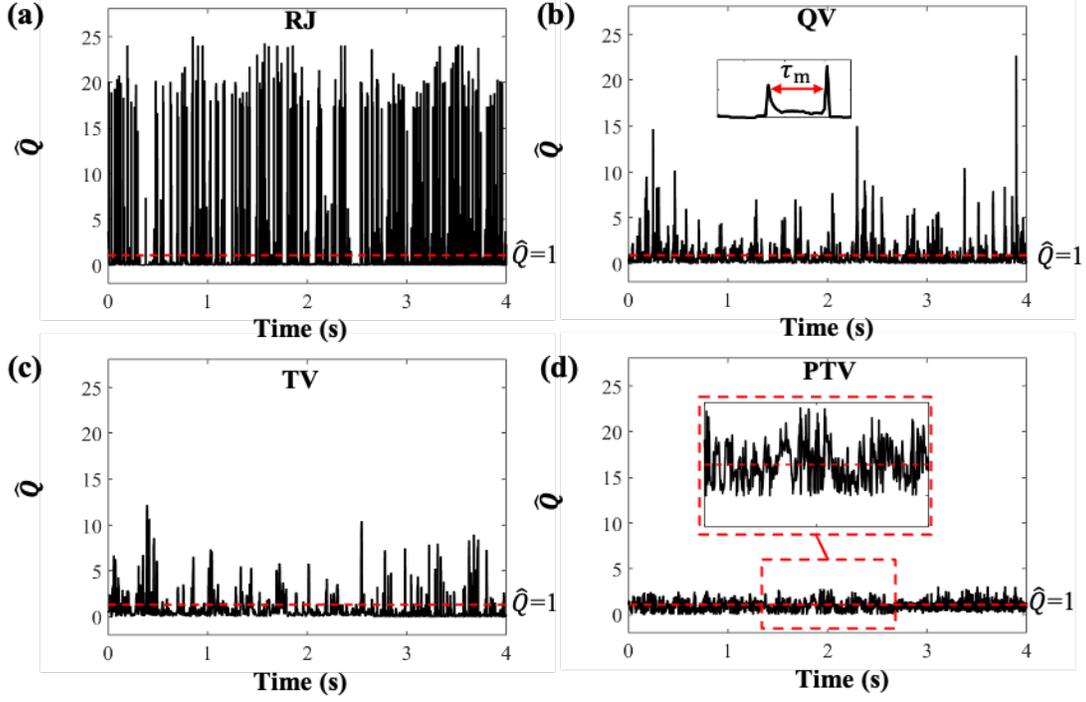

Figure 4. Comparison of instantaneous gas leakage ($\hat{Q}$) across different cavity closures. (a) Re-entrant jet (RJ) closure cavity with ventilation rate $Q_{AS} = 0.5$ SLPM and gas rate coefficient $C_{Qs} = 0.027$; (b) quad vortex (QV) closure cavity with $Q_{AS} = 0.6$ SLPM and $C_{Qs} = 0.033$; (c) twin vortex (TV) closure cavity with $Q_{AS} = 1.0$ SLPM and $C_{Qs} = 0.056$ and (d) pulsating twin vortex (PTV) closure cavity with $Q_{AS} = 7.0$ SLPM and $C_{Qs} = 0.39$. All the experiments are conducted under tunnel speed at $U_\infty = 3.0$ m/s and Froude number at $Fr = 9.6$. The inset figure in (b) depicts the definition of $\tau_m$ (peak interval, the time interval between adjacent local maxima of $\hat{Q}$) to quantify the temporal occurrence of the gas leakage fluctuations and the inset figure in (d) is used to showcase more frequent gas leakage spikes in PTV case in comparison to other cases.

The gas leakage magnitude under different closure types is further quantified by the histograms of normalized gas leakage ($\hat{Q}$) as shown in Fig. 5. Fig. 5 represents the distribution of instantaneous normalized gas leakage over the whole measurement span for different closure types. Compared to the temporal variation of gas leakage illustrated in Fig. 4, Fig. 5 provides a more quantitive assessment of the variability of gas leakage for different closure types. For the supercavities with RJ, QV and TV closures, the $\hat{Q}$ has the highest probability at around average gas leakage rate (i.e., $\hat{Q} = 1$) and its probability decreases exponentially with increasing $\hat{Q}$ (Fig. 5a to c). Noteworthily, for RJ closure (Fig. 5a), $\hat{Q}$ histogram yields a local maximum at higher values (i.e., $\hat{Q} = 18$) while such local peak does not appear in the histograms of QV and TV (Fig. 5b and c). In comparison to other supercavity closure types, the $\hat{Q}$ distribution of PTV cavity has a significantly narrower range (i.e., from 0 to 2.8, Fig. 5d). To quantify the fluctuation of gas leakage with respect to the ventilation input under different flow and ventilation conditions, we define the instantaneous excessive gas leakage ($Q_E$, the amount of gas leakage higher than ventilation input) as Eqn. 4.

$$Q_E = Q_{GL} - Q_{AS}, \text{ when } Q_{GL} > Q_{AS}. \tag{4}$$

And the relative amount of excessive gas leakage over the whole measurement span ($\hat{Q}_E$) for each case can be defined as Eqn. 5.

$$\hat{Q}_E = \int_{Q_{AS}}^{\infty} Q_E/Q_{AS} P(Q_E) \, dQ_E \tag{5}$$

Accordingly, as the cavity transitions from RJ, to QV, to TV, and to PTV with increasing ventilation, the $\hat{Q}_E$ value drops monotonically from 5.28 (RJ), to 2.46 (QV), 2.43 (TV), and to 1.48 (PTV). Note that $\hat{Q}_E$ is non-dimensional and the decrease of $\hat{Q}_E$ does not imply the drop of fluctuation in the absolute values of gas leakage, rather a decline of the fluctuation of gas leakage relative to the ventilation input.

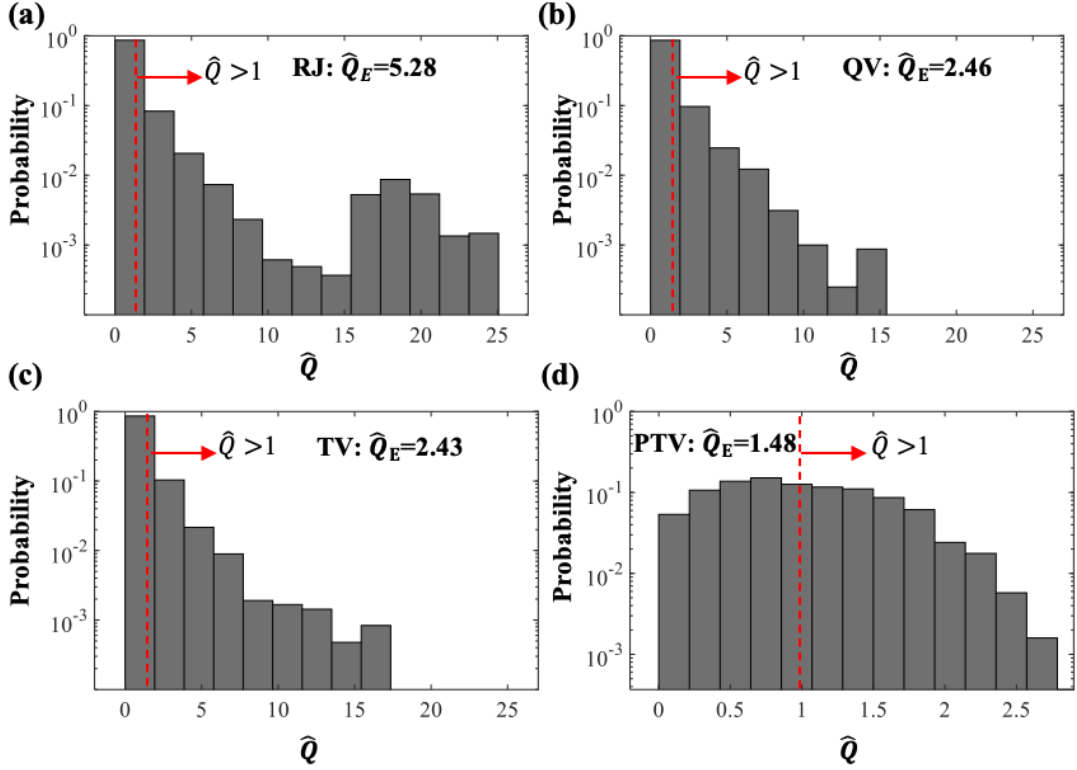

Figure 5. Histograms of instantaneous gas leakage rate ($\hat{Q}$) for (a) Re-entrant jet (RJ) closure cavity with ventilation rate $Q_{AS} = 0.5$ SLPM and gas rate coefficient $C_{Qs} = 0.027$; (b) Quad vortex (QV) closure cavity with $Q_{AS} = 0.6$ SLPM and $C_{Qs} = 0.033$; (c) Twin vortex (TV) closure cavity with $Q_{AS} = 2.0$ SLPM and $C_{Qs} = 0.056$ and (d) Pulsating twin vortex (PTV) closure cavity with $Q_{AS} = 7.0$ SLPM and $C_{Qs} = 0.39$. All the experiments are conducted under tunnel speed at $U_\infty = 3.0$ m/s and Froude number at $Fr = 9.6$. The excessive gas leakage $\hat{Q}_E$ corresponding to each case is also labeled in the figure.

The temporal characteristics of gas leakage fluctuations under different closure conditions is investigated through the statistical distribution of non-dimensionalized time interval between adjacent local maxima of gas leakage signals (illustrated using the inset figure of Fig. 4b), referred to as local Strouhal number ($St_m$) defined below:

$$St_m = \frac{(1/\tau_m)d_C}{U_\infty} \quad (6)$$

Fig. 6 shows the histograms of $St_m$ corresponding to different closure conditions presented in Fig. 5. As the figure shows, the histograms of RJ, QV and TV yield a maximum $St_m$ near 0.10. Particularly, the $St_m$ for the vortex-based closures (QV and TV) shows clear log-normal distributions with the peak $St_m$ increases slightly when transitioning from QV to TV. Remarkably, in contrast to QV and TV, the histogram of $St_m$ for RJ (Fig. 6a) exhibits a secondary peak at around $St_m = 0.35$, corresponding to the second mode of the gas leakage presented in the $\hat{Q}$ histogram of RJ (Fig.5a). As for the PTV, the peak of $St_m$ distribution increases to about 0.22 and broadens in comparison to those in RJ, QV, and TV cases.

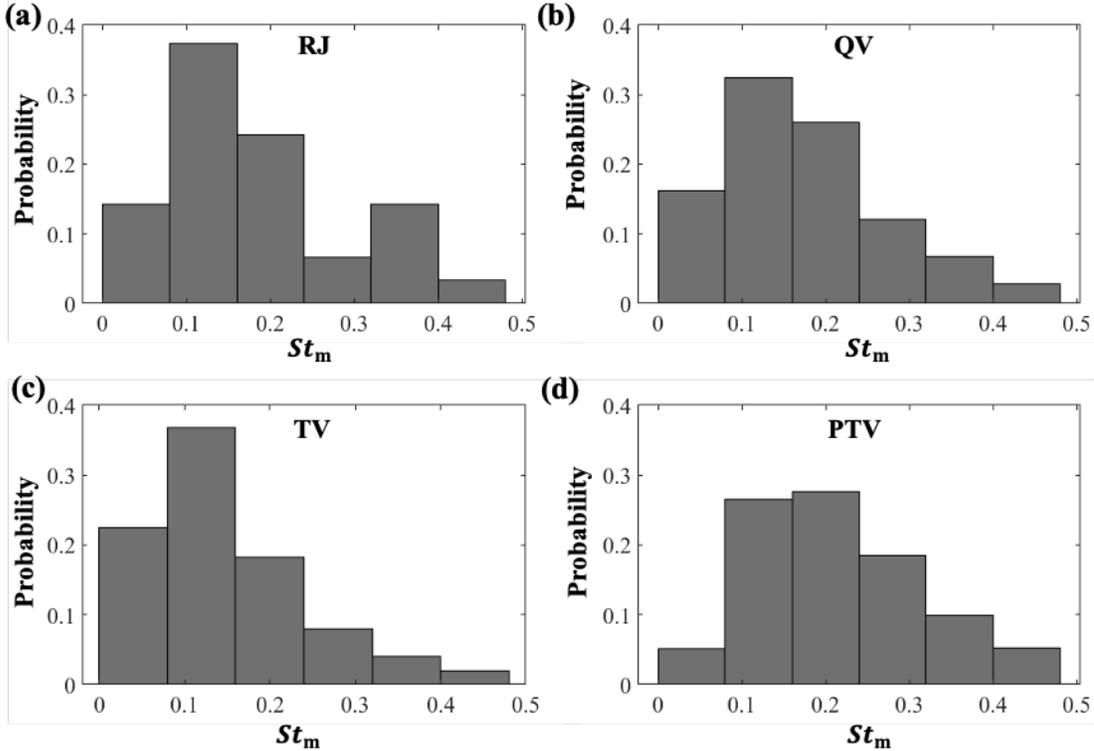

Figure 6. Histograms of local Strouhal number ($St_m$) of gas leakage peaks for (a) Re-entrant jet (RJ) closure cavity with ventilation rate $Q_{AS} = 0.5$ SLPM and gas rate coefficient $C_{Qs} = 0.027$; (b) quad vortex (QV) closure cavity with $Q_{AS} = 0.6$ SLPM and $C_{Qs} = 0.033$; (c) twin vortex (TV) closure cavity with $Q_{AS} = 2.0$ SLPM and $C_{Qs} = 0.056$ and (d) pulsating twin vortex (PTV) closure cavity with $Q_{AS} = 7.0$ SLPM and $C_{Qs} = 0.39$. All the experiments are conducted under tunnel speed at $U_\infty = 3.0$ m/s and Froude number at $Fr = 9.6$.

We attribute the above trends in $\hat{Q}$ and $St_m$ to the presence of different gas leakage modes and the variation in their dominance across different closure conditions. Specifically, as pointed out by Wu et al. (2019), under RJ closure, the gas leakage occurs through both re-entrant jet impingement ($Q_{RJ}$) and vortex tube gas leakage ($Q_{VT}$) (Fig. 7a). These two mechanisms contribute to the two gas leakage modes observed in the corresponding $\hat{Q}$ and $St_m$ histograms (i.e, Fig.5a and 6a), respectively. In more detail, the $Q_{RJ}$ is caused by the toroidal vortex ejected into the water from the cavity triggered by the water jet impinging the liquid-gas interface of the cavity. The fluctuation in $Q_{RJ}$ can be associated with the frequency of re-entrant jet impingement to the cavity surface near closure region. Using high-speed imaging, under the same flow condition as the RJ cavity discussed previously, we can estimate the frequency of periodical entrance of water jet to the cavity to be around 95 Hz (corresponding Strouhal number of 0.32), coincident with the $St_m$ peak around 0.35. In comparison, the $Q_{VT}$ is related to the vortex tubes developed behind the RJ cavity, analogous to the formation of shedding vortex behind an aircraft wing (Wu et al. 2019). We suggest that the fluctuation in $Q_{VT}$ is influenced by the instability of vortex tubes. According to Leweke et al. (2016), two types of instabilities can occur for a pair of vortex tubes shed from a cavity, i.e., Crow instability and elliptic instability, both of which are observed in our current experiments as shown in Fig. 8(a). The Crow instability corresponds to the long wavelength oscillation of the vortex tubes and is manifested as the repeated arc shape bubbles downstream of the cavity (Fig. 8a), while the elliptic instability induces short wavelength oscillation of vortex tubes and triggers the breakup of tube filaments to small bubbles (Fig. 8b). Under the Reynolds number in our range of measurement (Re = $3.3 \times 10^4$), the characteristic wavelength of the Crow instability for a vortex tube in a pair of counter-rotating vortex tubes is approximately 16 times of the vortex tube diameter (Leweke et al. 2016). The high-speed visualization of supercavity closure reveals that the diameter of the vortex tube for the RJ case discussed in this section is around 1.0 cm corresponding to the characteristic wavelength of Crow instability around 16.0 cm for the present case. Combining the tunnel speed of 3 m/s, the characteristic frequency of Crow instability caused vortex tube instability can be estimated to be 19 Hz (i.e., $f = U_\infty/\lambda = 19$ Hz, corresponding to Strouhal number of 0.06). This Strouhal number value is generally in accordance with the peak value observed from the bubble data for the RJ case (0.10, Fig. 6a) although the latter has a higher peak value with a broader/broadened distribution. We suggest that such discrepancies are due to the $St_m$ distribution observed from bubble volume change data is a result from the Crow instability caused by multiple vortex tubes and other type of vortex instabilities including elliptic instability.

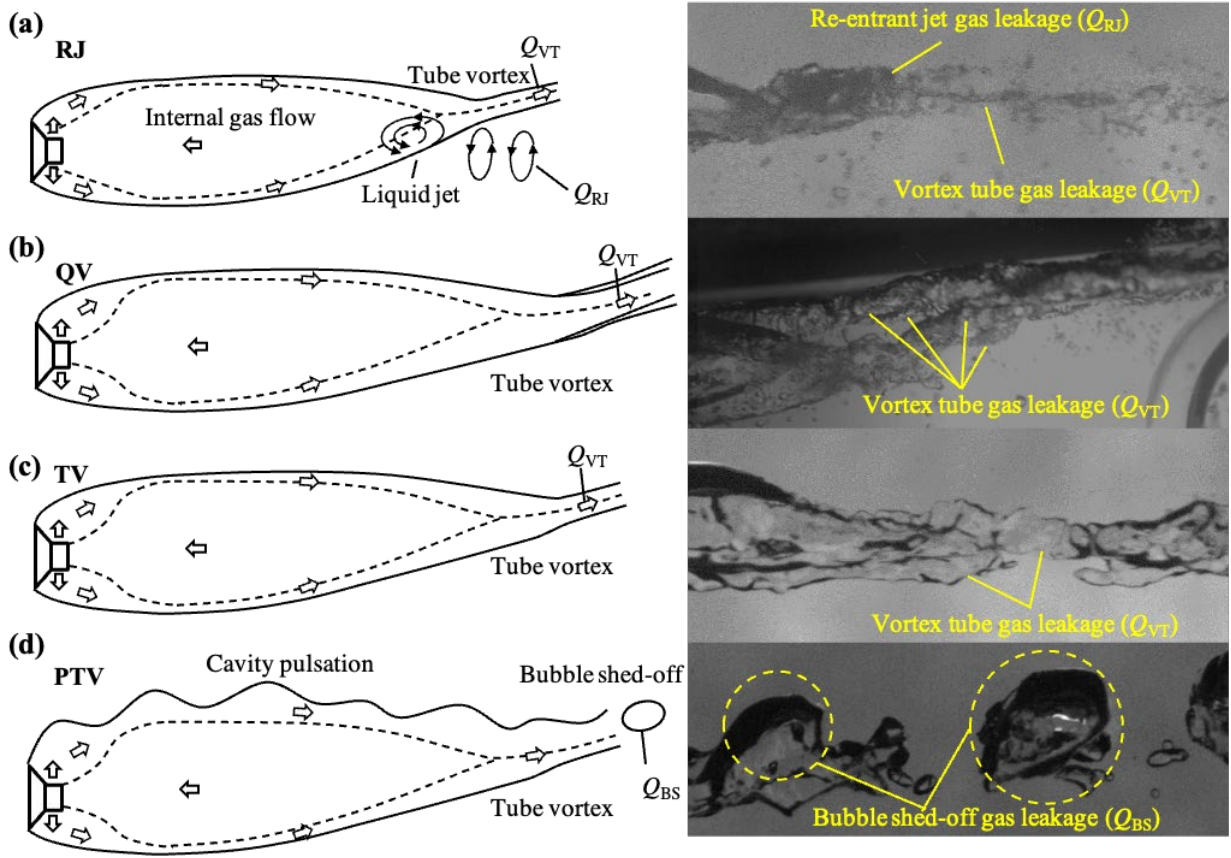

Figure 7. The schematics and the corresponding high-speed images showing the gas leakage mechanism for supercavity under (a) re-entrant jet closure, (b) quad vortex closure, (c) twin vortex closure, and (d) pulsating twin vortex closure. Note that the image showing the quad vortex closure is captured in a tilt angle to better illustrate the presence of four vortex tubes extended from the supercavity.

As the cavity transition to vortex-based closures, the re-entrant jet gas leakage diminishes according to Wu et al. (2019) and vortex tube leakage becomes the dominant leakage mechanism (Fig. 7b and c). Correspondingly, the local maximum in the $\hat{Q}$ histogram ($\hat{Q} = 18$) of the RJ cavity (Fig. 5a) disappears in those of QV (Fig. 5b) and TV cavities (Fig. 5c), and so does the higher frequency peak in the $St_m$ histogram (Fig. 6a). The $St_m$ distributions of both QV (Fig. 6b) and TV cavities (Fig. 6c) yield only one peak near the Strouhal number associated with vortex tube instabilities (Fig. 6b and c). Remarkably, in comparison to that of TV, the $St_m$ histogram of QV cavity appears to be more broadened (i.e., distributed more evenly over a broader range of frequencies) potentially due to the presence of more vortex filaments under QV closure.

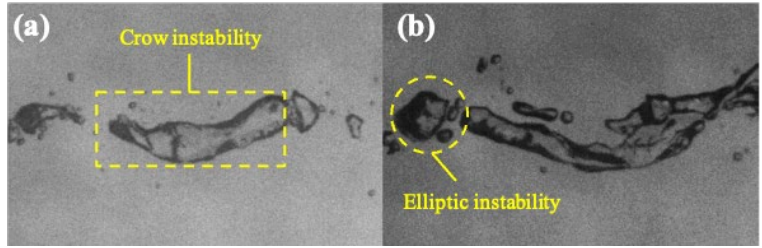

Figure 8. Different types of vortex tube instability observed in our experiments. (a) The Crow instability of vortex tubes under re-entrant jet closure (labeled by the dashed rectangle) and (b) the elliptic instability of vortex tubes under quad vortex closure (labeled by the dashed circle).

With further increase in ventilation, the cavity transitions to PTV cavity and starts to pulsate, causing the shed-off of the large bubble pockets (Fig. 7d, Michel 1984). This phenomenon has been noted in the literature (Michel 1984) and we denote the gas leakage associated with this pulsation-induced bubble shed-off as $Q_{BS}$. According to Skidmore (2016), the cavity tends to pulsate at its natural frequency determined by the Eqn. 6 (Song 1961). In this equation, the $\gamma$ is the specific heat of the ventilation gas, $S$ is the surface area of the cavity, and the $D_C$ is the cavity maximum diameter (the other symbols have been introduced in the previous sections).

$$f_0 = \frac{1}{2\pi} \sqrt{\frac{2\pi\gamma P_C}{6\rho_w S \ln\frac{d_C}{D_C}}} \qquad (7)$$

Through Eqn. 7, the supercavity natural frequency is determined to be 64 Hz (corresponding to a Strouhal number of 0.21) with the $S$ estimated using the circular fitted cavity cross-sections from side view of the cavity digital images. This value is in a good agreement with the averaged $St_m$ of about 0.2 for PTV cavity (Fig. 6d), providing a strong support to the connection between the gas leakage fluctuation and bubble shed-off mechanisms discussed above.

### 3.2. Effect of tunnel speed on supercavity gas leakage

In this section, we will investigate the effects of tunnel speed on temporal characteristics of gas leakage under three closure types, i.e., reentrant-jet (RJ), twin vortex (TV), and pulsating twin vortex (PTV). Note that quad vortex (QV) cavity described in the previous section is not included in the section because of the lack of QV data under different tunnel speeds. In addition, for each closure type, the comparison is presented for different tunnel speeds with fixed ventilation rate instead of using dimensionless numbers (i.e., $Fr$ and $C_{Qs}$) as those in the previous section since these dimensionless numbers are only suitable for characterizing the overall geometry of the supercavity not detailed gas leakage physics at the cavity closure (Karn et al. 2016a). For a RJ supercavity, with increasing tunnel speed under the fixed ventilation rate, the excessive gas leakage ($\widehat{Q}_E$) first decreases and then plateaus (Fig. 9a), and $\widehat{Q}$ distribution narrows and its local maximum (at $\widehat{Q} \approx 18$) gradually disappears (comparing Fig. 9b and Fig. 9c). Accordingly, averaged local Strouhal number ($\overline{St_m}$) associated with gas leakage fluctuation increases (Fig. 10a) and the secondary peak in the $St_m$ distribution drops out and the entire distribution shifts to higher values/frequencies (Fig. 10b and c). These trends can be explained by the growing dominance of reentrant-jet gas leakage ($Q_{RJ}$) over twin vortex gas leakage ($Q_{TV}$) with increasing tunnel speed, which suppresses the secondary mode associated with $Q_{TV}$ in $\hat{Q}$ and $St_m$ distributions. Correspondingly, the suppression of the secondary leakage mode reduces the excessive gas leakage with respect to ventilation input and thus $\hat{Q}_E$. Moreover, according to Le et al. (1993), with increasing tunnel speed under fixed ventilation, the liquid re-entrant jet enters the cavity more repeatedly, leading to the increasing frequency in gas leakage fluctuation and the corresponding shift of $St_m$ to higher values.

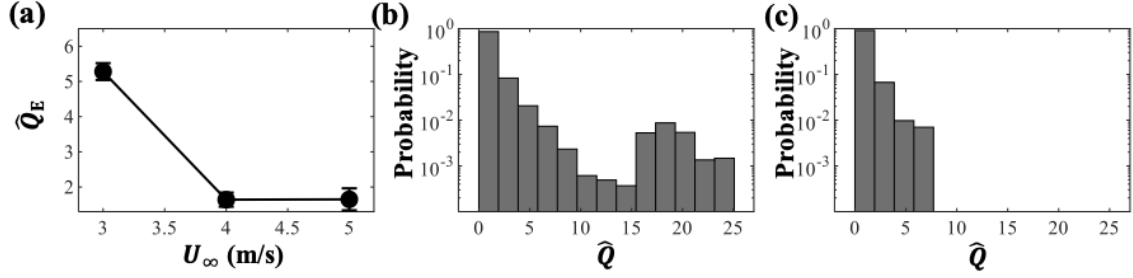

Figure 9. (a) Excessive gas leakage ($\hat{Q}_E$) under different tunnel speeds ($U_\infty$) and the histograms of the instantaneous gas leakage corresponding to (b) $U_\infty = 3.0$ m/s and (c) $U_\infty = 5.0$ m/s for re-entrant jet (RJ) closure supercavity. The ventilation rate is set at constant value of 0.5 SLPM.

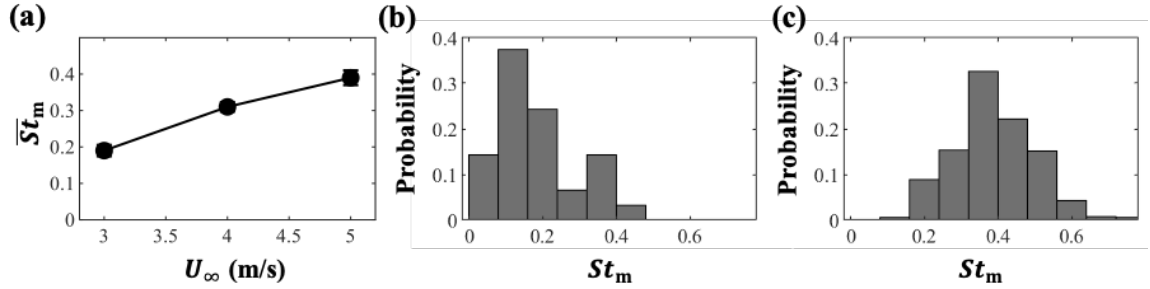

Figure 10. (a) Averaged gas leakage local Strouhal number ($\bar{St}_m$) under different tunnel speeds ($U_\infty$) and the histograms of the local Strouhal number ($St_m$) corresponding to (b) $U_\infty = 3.0$ m/s and (c) $U_\infty = 5.0$ m/s for re-entrant jet (RJ) closure supercavity. The ventilation rate is set at constant value of 0.5 SLPM.

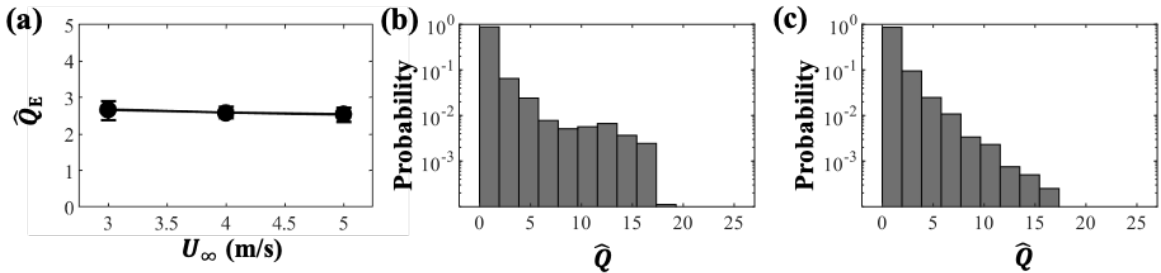

Figure 11. (a) Excessive gas leakage ($\hat{Q}_E$) under different tunnel speeds ($U_\infty$) and the histograms of the instantaneous gas leakage corresponding to (b) $U_\infty = 3.0$ m/s and (c) $U_\infty = 5.0$ m/s for twin vortex (TV) closure supercavity. The ventilation rate is set at constant value of 2.0 SLPM.

For a TV supercavity, as shown in Fig. 11, the increasing tunnel speed with fixed ventilation leads to a slight decrease in $\hat{Q}_E$ (Fig. 11a) and a reduction in the probability of high values in $\hat{Q}$ distribution (Fig. 11b and c). Correspondingly, the $\bar{St}_m$ increases (Fig. 12a) and the $St_m$ distribution shifts to the right (Fig. 12b and c) with increasing tunnel speed. Under TV closure, the gas leakage is dominated by $Q_{TV}$. The increasing flow speed can result in more frequent vortex tube break up according to Leweke et al. (2016) and more dominant $Q_{TV}$ (i.e., the $Q_{BS}$ mode at lower $U_\infty$ is suppressed). Such a phenomenon causes more frequent gas leakage fluctuation, the right shift of $St_m$ distribution, and slight lowering of $\hat{Q}_E$ following the same reasoning used to explain the trends for RJ supercavity mentioned above. It is worth noting that TV supercavity tends to transition to PTV with decreasing tunnel speed. This tendency of transition may explain the increase of the probability of high $\hat{Q}$ in the $\hat{Q}$ distribution (i.e., the small bump at $\hat{Q} \approx 13$ in Fig. 11b) at lower tunnel speed.

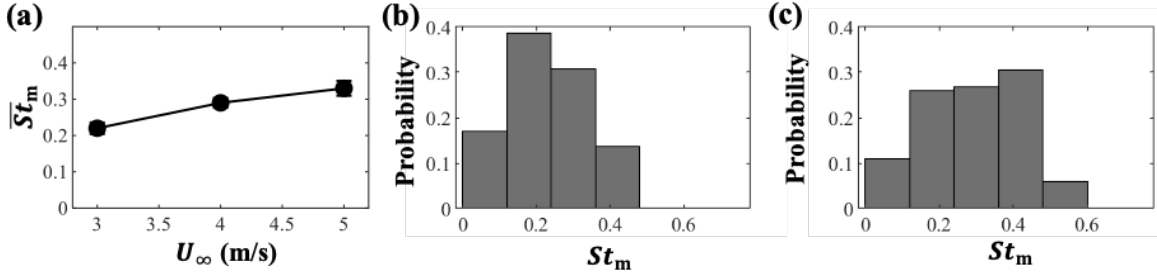

Figure 12. (a) Averaged gas leakage local Strouhal number ($\bar{St}_m$) under different tunnel speeds ($U_\infty$) and the histograms of the local Strouhal number ($St_m$) corresponding to (b) $U_\infty = 3.0$ m/s and (c) $U_\infty = 5.0$ m/s for twin vortex (TV) closure supercavity. The ventilation rate is set at constant value of 2.0 SLPM.

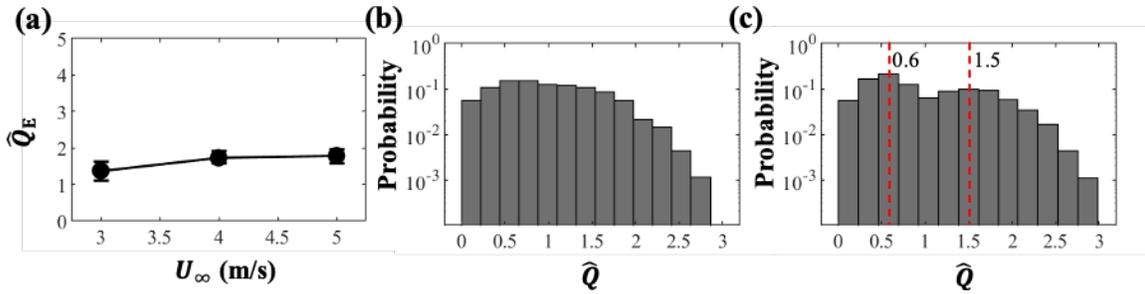

Figure 13. (a) Excessive gas leakage ($\hat{Q}_E$) under different tunnel speeds ($U_\infty$) and the histograms of the instantaneous gas leakage corresponding to (b) $U_\infty = 3.0$ m/s and (c) $U_\infty = 5.0$ m/s for pulsating twin vortex (PTV) closure supercavity. Note that the two peaks of instantaneous gas leakage distribution are marked using red dashed vertical lines in (c). The ventilation rate is set at constant value of 9.0 SLPM.

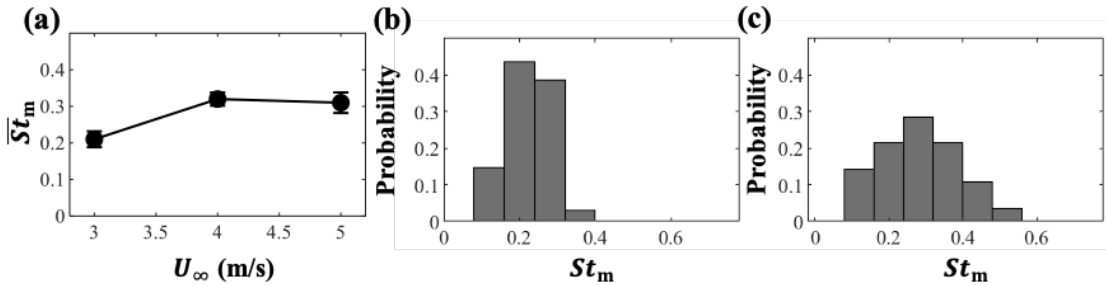

Figure 14. (a) Averaged gas leakage local Strouhal number ($\bar{St}_m$) under different tunnel speeds ($U_\infty$) and the histograms of the local Strouhal number ($St_m$) corresponding to (b) $U_\infty = 3.0$ m/s and (c) $U_\infty = 5.0$ m/s for pulsating twin vortex (PTV) closure supercavity. The ventilation rate is set at constant value of 9.0 SLPM.

For a PTV supercavity, it is observed that the increasing tunnel speed under fixed ventilation results in a small increase of $\hat{Q}_E$ (Fig. 13a) and the $\hat{Q}$ exhibits a double-peak distribution under high $U_\infty$ (Fig. 13c) instead of a single peak mode for lower $U_\infty$ (Fig. 13b). Accordingly, the $\bar{St}_m$ increases then plateaus (Fig. 14a) and the $St_m$ distribution becomes broader and shifts to the right with increasing $U_\infty$ (Fig. 14 b and c). Under PTV closure, the gas leakage is dominated by $Q_{BS}$. However, with increasing tunnel speed under fixed ventilation, the supercavity gradually transitions to TV closure and the gas leakage becomes more influenced by $Q_{VT}$, leading to the emergence of second mode in the $\hat{Q}$ distribution (Fig. 13c) and an increase of $\hat{Q}_E$. Correspondingly, the interplay between the pulsation-induced bubble shed-off and vortex tube breakup causes the broadening and right shift of $St_m$ distribution (Fig. 14b and c).

### 3.3. Effect of ventilation rate on supercavity gas leakage

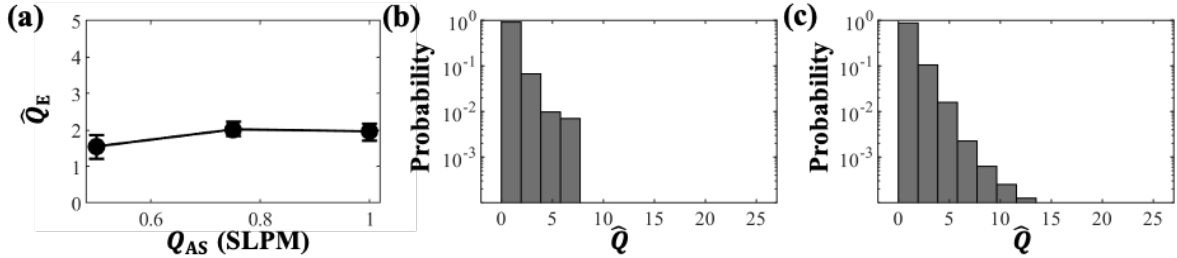

Figure 15. (a) Excessive gas leakage ($\hat{Q}_E$) under different ventilation rates ($Q_{AS}$) and the histograms of the instantaneous gas leakage corresponding to (b) $Q_{AS} = 0.5$ SLPM and (c) $Q_{AS} = 1.0$ SLPM for re-entrant jet (RJ) closure supercavity. The tunnel speed is set at constant value of 5.0 m/s.

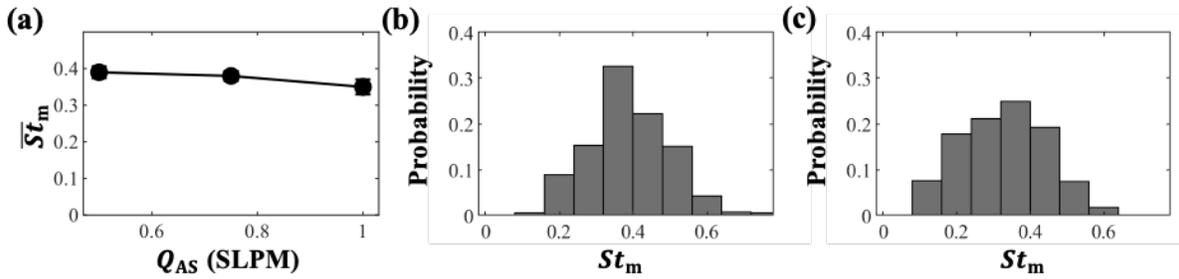

Figure 16. (a) Averaged gas leakage local Strouhal number ($\overline{St}_m$) under different ventilation rates ($Q_{AS}$) and the histograms of the local Strouhal number ($St_m$) corresponding to (b) $Q_{AS} = 0.5$ SLPM and (c) $Q_{AS} = 1.0$ SLPM for re-entrant jet (RJ) closure supercavity. The tunnel speed is set at constant value of 5.0 m/s.

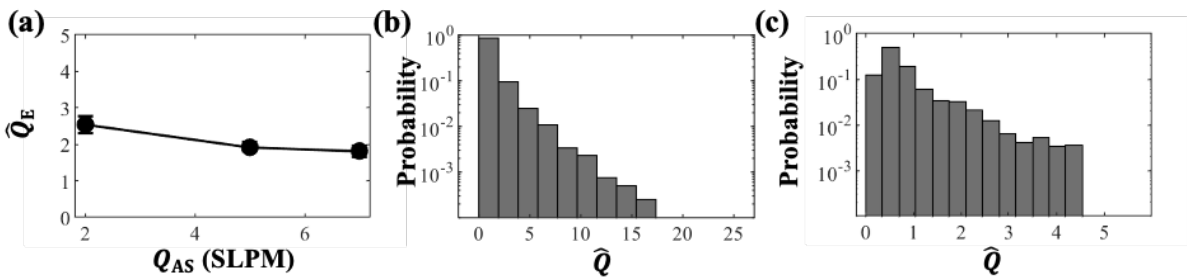

Figure 17. (a) Excessive gas leakage ($\hat{Q}_E$) under different ventilation rates ($Q_{AS}$) and the histograms of the instantaneous gas leakage corresponding to (b) $Q_{AS} = 2.0$ SLPM and (c) $Q_{AS} = 7.0$ SLPM for twin vortex (TV) closure supercavity. The tunnel speed is set at constant value of 5.0 m/s.

The comparison of supercavity gas leakage under different ventilation rates ($Q_{AS}$) is conducted under a fixed tunnel speed of 5.0 m/s. This tunnel speed is chosen since it offers the most abundant cases with different ventilation rates in our experiments. For a RJ supercavity, the increasing ventilation under fixed tunnel speed leads to a small increase of $\hat{Q}_E$ (Fig. 15a) and an increase in the probability of high values in $\hat{Q}$ distribution (Fig. 15b and c). Correspondingly, the $\overline{St}_m$ drops slightly (Fig. 16a) and the $St_m$ distribution broadens with peak value (at $St_m \approx 0.35$) lowers (Fig. 16b and c) with increasing $Q_{AS}$. Following the general physical framework laid out in Section 3.1, these trends can be explained by the kicking in of vortex tube gas leakage ($Q_{VT}$) as the $Q_{AS}$ increases, which increases $\hat{Q}_E$ and causes the change in the distributions of $\hat{Q}$ and $St_m$.

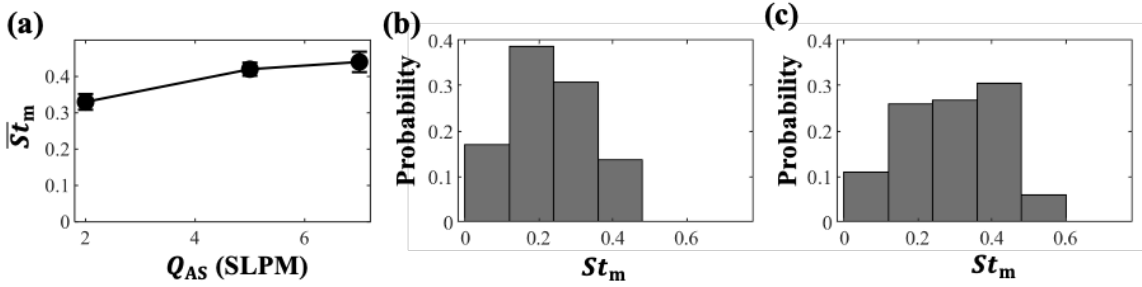

Figure 18. (a) Averaged gas leakage local Strouhal number ($\bar{St}_m$) under different ventilation rates ($Q_{AS}$) and the histograms of the local Strouhal number ($St_m$) corresponding to (b) $Q_{AS} = 2.0$ SLPM and (c) $Q_{AS} = 7.0$ SLPM for twin vortex (TV) closure supercavity. The tunnel speed is set at constant value of 5.0 m/s.

For a TV supercavity, with increasing ventilation, the $\hat{Q}_E$ drops (Fig. 17a) and the distribution of $\hat{Q}$ narrows (Fig. 17b and c), while the $\bar{St}_m$ increases (Fig. 18a) and the $St_m$ distribution shifts towards high values with a broadened peak (Fig.18b and c). Since $Q_{VT}$ is the only dominant leakage mode under TV closure, the increasing ventilation can lead to faster internal gas flow inside vortex tubes and induces vortex tube breakup at higher frequency and shorter wavelength (Leweke et al. 2016). As the results, the gas leakage fluctuations caused by vortex breakup decreases substantially compared to ventilation input, leading to narrower $\hat{Q}$ distribution, decreasing $\hat{Q}_E$, and right shift and broadening of the $St_m$ with increasing $\bar{St}_m$.

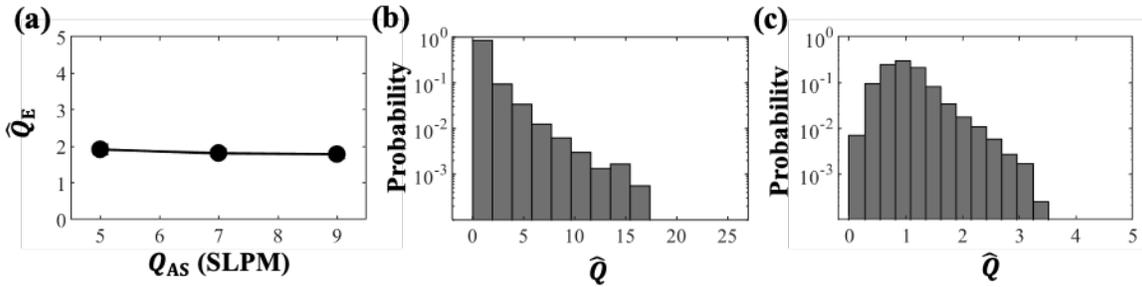

Figure 19. (a) Excessive gas leakage ($\hat{Q}_E$) under different ventilation rates ($Q_{AS}$) and the histograms of the instantaneous gas leakage corresponding to (b) $Q_{AS} = 5.0$ SLPM and (c) $Q_{AS} = 9.0$ SLPM for pulsating twin vortex (PTV) closure supercavity. The tunnel speed is set at constant value of 3.0 m/s.

For a PTV supercavity, the increasing ventilation yields little influence on $\hat{Q}_E$ (Fig. 19a) but results in significant narrowing of $\hat{Q}$ distribution (Fig. 19b and c). Accordingly, the $\bar{St}_m$ increases slightly (Fig. 20a) and the $St_m$ distribution exhibits a more prominent peak around $St_m = 0.2$ at higher ventilation (Fig. 20c) in comparison to the lower ventilation case (Fig. 20b). Such trends are the results of growing dominance of the gas leakage associated with cavity pulsation ($Q_{BS}$) over that caused by vortex tube instability ($Q_{VT}$), which causes gas leakage fluctuation to match more and more the natural frequency of the cavity pulsation. Noteworthily, despite the change of $St_m$ distribution, the peak probability of $St_m$ remains around 0.2 across the ventilation conditions tested in our experiments (Fig. 6d, Fig. 20b and c). As shown in the literature (Fronzeo et al. 2019, Xu et al. 2021), the cavity geometry and pressure distribution upstream of the closure region (i.e., where vortex tubes start) remains unchanged with increasing ventilation after the supercavity transitions to vortex-based closure. Therefore, according to Eqn. 7, the cavity natural frequency remains constant with increasing ventilation and leads to relative invariant of $St_m$ peak across the range of ventilation investigated in our experiments.

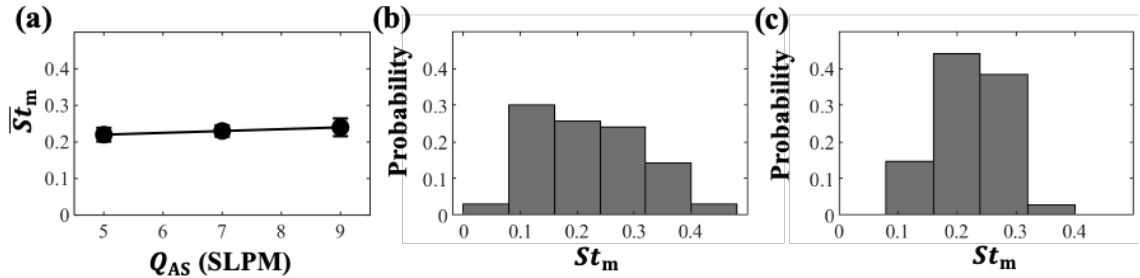

Figure 20. (a) Averaged gas leakage local Strouhal number ($\overline{St}_m$) under different ventilation rates ($Q_{AS}$) and the histograms of the local Strouhal number ($St_m$) corresponding to (b) $Q_{AS}$ = 5.0 SLPM and (c) $Q_{AS}$ = 9.0 SLPM for pulsating twin vortex (PTV) closure supercavity. The tunnel speed is set at constant value of 3.0 m/s.

### 3.4. Supercavity controllability based on gas leakage characteristics

In this section, we focus on assessing the controllability of a supercavity under different flow, ventilation, and closure conditions from the standpoint of minimizing the instantaneous ventilation deficit when the cavity experiences gas leakage fluctuations. As suggested in Sun et al. (2019), excessive (i.e., higher than ventilation input) gas leakage can disrupt cavity internal flow and geometry and undermine cavity stability. Accordingly, accurate prediction of the amplitude and occurrence of excessive gas leakage will allow us to control the stability of supercavity using appropriate ventilation compensation. Therefore, we introduce two metrics to characterize the controllability of a supercavity based on the change of magnitude of excessive gas leakage and the predictability of its occurrence.

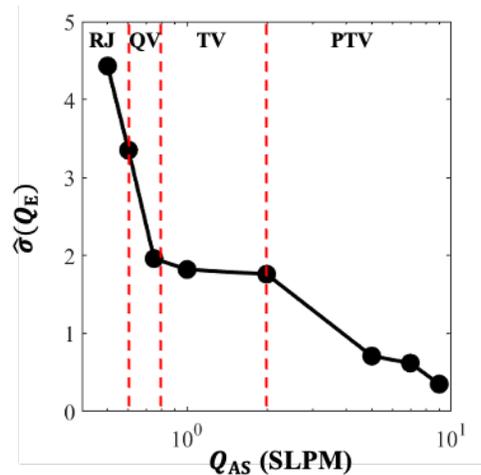

Figure 21. The dependence of normalized standard deviation of excessive gas leakage ($\hat{\sigma}(Q_E)$) upon the changing ventilation ($Q_{AS}$) under a fixed tunnel speed ($U_\infty$ = 3.0 m/s).

First, we use the normalized standard deviation of excessive gas leakage (Eqn. 8, where $Q_E$ corresponds to the instantaneous gas leakage higher than $Q_{AS}$, and $\sigma(Q_E)$ is the standard deviation of $Q_E$) to quantify the relative change of ventilation input needed to compensate the extra of gas loss to sustain a cavity. We expect lower $\hat{\sigma}(Q_E)$ implies better controllability of a ventilated supercavity. As shown in the Fig. 21, with increasing ventilation, the $\hat{\sigma}(Q_\text{E})$ shows monotonic decrease first during the transition from a RJ cavity to a QV cavity. The $\hat{\sigma}(Q_\text{E})$ value exhibits a plateau region once cavity reaches a TV closure, indicating that the change of ventilation does not significantly affect the fluctuation of gas leakage of the supercavity in this regime. As the cavity transition to PTV, the $\hat{\sigma}(Q_\text{E})$ starts to drop again at a gentler slope in comparison to that in the RJ and QV regimes.

$$\hat{\sigma}(Q_\text{E}) = \frac{\sigma(Q_\text{E})}{Q_\text{AS}} \tag{8}$$

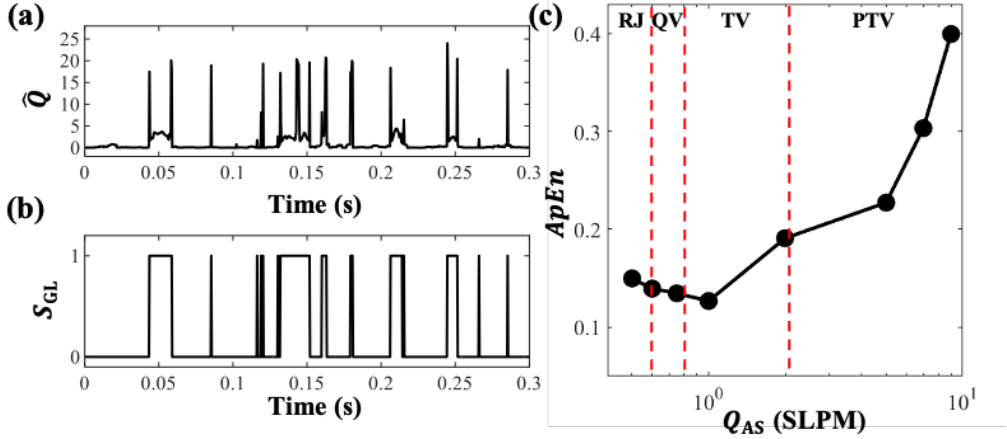

Figure 22. (a) A sample of original gas leakage signal under the tunnel speed at $U_\infty = 3.0$ m/s and ventilation rate at $Q_\text{AS} = 0.5$ SLPM and (b) the corresponding simplified controlling signal $S_\text{GL}$. (b) The dependence of gas approximate entropy ($ApEn$) upon the changing ventilation ($Q_\text{AS}$) under the fixed tunnel speed at $U_\infty = 3.0$ m/s.

The second metric is the approximate entropy ($ApEn$) of the simplified gas leakage signal ($S_\text{GL}$) which is used to quantify the regularity and predictability of the occurrence of excessive gas leakage. The $ApEn$ was introduced in Pincus (1991) as a way to provide quantitative characterization of similarity among the subsections of a time series signal. The higher $ApEn$ indicates more irregular and unpredictable time series signals. It has been commonly used for analysis of time series signals including physiological signal like respiratory signals (Chen et al. 2005), predicting machine failures (Yan & Gao 2007), and financial data (Pincus 2008). The $ApEn$ of a signal is usually calculated using an iterative approach (Yan & Gao 2007). The time series signal with a length of $n$ is first divided to subsections using prescribed lengths $m$. Subsequently, the distances between the signal subsections are quantified by the averaged Euclidean distance between the consecutive sections divided by $m$. The $ApEn$ is calculated as the minimal signal section distance with respective to different $m$ divided by $n - m + 1$. The second division of the calculation is used to account for the required observation length of time series signals to make a prediction of following signal patterns. It can be summarized that it is more difficult to find the similar sections within the signal for the time series signal with high $ApEn$, and it requires longer observation of the signal to make a prediction of subsequent signal patterns. The $S_\text{GL}$ is defined using Eqn. 9.

$$S_\text{GL} = \begin{cases} 1, \hat{Q} > 1 \\ 0, else \end{cases} \tag{9}$$

As illustrated in Fig. 22(a) and (b), the $S_\text{GL}$ transforms gas leakage signals into a square-wave TTL signal with "on" state corresponding to the occurrence of excessive gas leakage events. Note that gas leakage is below the ventilation input and there is no need to have extra ventilation input to compensate excessive gas leakage. The $ApEn$ of $S_{GL}$ is calculated using the method provided by Yan & Gao (2007). As shown in Fig. 22(c), with increasing ventilation under the same tunnel speed, the $ApEn$ first decreases as the cavity transitions from RJ to TV closure under moderate ventilation rate. Once the $Q_{AS}$ becomes larger than 1.0 SLPM in our experiments (i.e., at the point when a TV cavity with high ventilation rate transitions to a PTV cavity), the $ApEn$ increases substantially with increasing ventilation, indicating more unpredictable and irregular patterns of gas leakage signals.

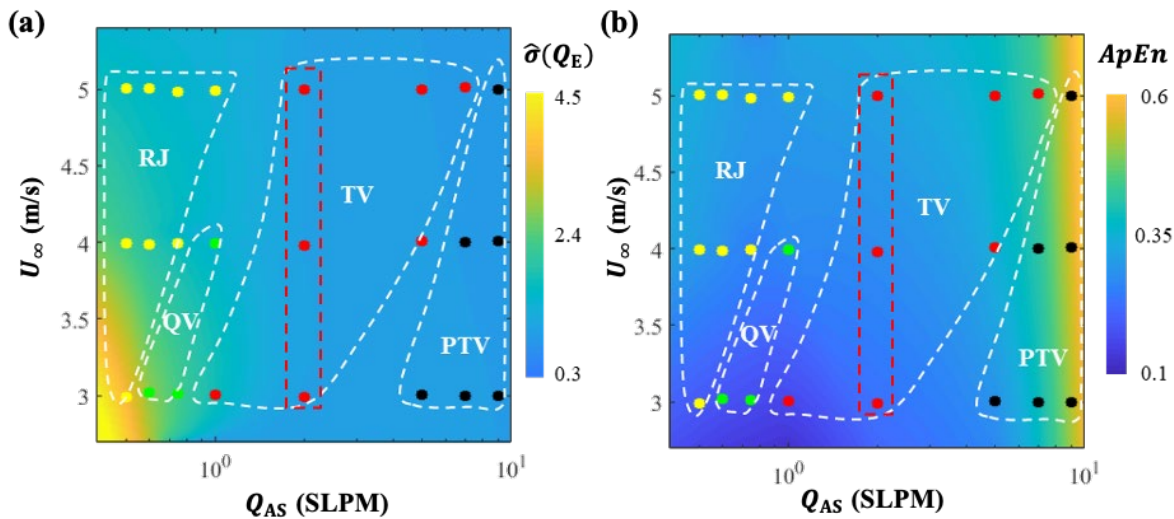

Figure 23. Maps of (a) normalized standard deviation of excessive gas leakage $\hat{\sigma}(Q_E)$ and (b) the approximate entropy $ApEn$ over the tunnel speed ($U_\infty$) and ventilation rate ($Q_{AS}$). The dots in the figure correspond to the experimental data points for the present study. The yellow, green, red, and black dots correspond to re-entrant jet (RJ), quad vortex (QV), twin vortex, and pulsating twin vortex (PTV) closures, respectively. The white dashed lines mark the approximate boundaries between different closure regimes. The possible regimes showing the optimal controllability (i.e., the values of both $\hat{\sigma}(Q_E)$ and $ApEn$ are relatively low) are highlighted by the red dashed rectangles.

Using the two defined metrics above, we can depict maps of $\hat{\sigma}(Q_E)$ and $ApEn$ upon the changing tunnel speed and ventilation rate under different closure conditions (Fig. 23). The regions marked by the white dashed lines only offer an approximate demarcation of supercavities with different closure types based on our current experimental data points. As abovementioned, the lower values (toward blue color) of $\hat{\sigma}(Q_E)$ and $ApEn$ imply easier ventilation adjustment to accommodate the occurrence of excessive gas leakage and better controllability of a ventilated supercavity. As shown in Fig. 23a, the region of high $\hat{\sigma}(Q_E)$ is situated in the lower range of $U_\infty$ and $Q_{AS}$, corresponding to the supercavity under RJ and QV closure. With the transition of the cavity to TV, the values of $\hat{\sigma}(Q_E)$ decrease and plateau, and then reduces slightly when the supercavity moves to the PTV regime. In contrast, for $ApEn$ map (Fig. 23b), the PTV regime yields the highest $ApEn$, and the lowest $ApEn$ is obtained in the lower range of $U_\infty$ and $Q_{AS}$ where the supercavity is under RJ and QV closure. Interestingly, at moderate $Q_{AS}$ under TV closure, the $ApEn$ is only slightly higher above the minimum and significantly lower than those under PTV and $\hat{\sigma}(Q_E)$ stays relatively unchanged within the range of $U_\infty$ and $Q_{AS}$ in our experiments. Therefore, considering both metrics (i.e., $\hat{\sigma}(Q_E)$ and $ApEn$), we expect that the supercavity operating under TV closure with moderate ventilation (i.e., the region highlighted using red dashed rectangle in Fig. 23) is optimal for ventilation-based controls. Specifically, within TV regime, the supecavity requires relatively low ventilation compensation to compensate the extra gas loss due to gas leakage fluctuation and $\hat{\sigma}(Q_E)$ is not affected significantly by the change of ventilation. In the meantime, the $ApEn$ remains low for the TV regime in comparison to PTV case, indicating a regular and more predictable gas leakage which potentially benefits the development of supercavity controlling strategy based on ventilation compensation.

## 4. Conclusions and Discussions

We conduct a systematic investigation of the gas leakage characteristics of a ventilated supercavity under different closure conditions including re-entrant jet (RJ), quad vortex (QV), twin vortex (TV), and pulsating twin vortex (PTV), generated from different tunnel speeds and ventilation conditions. Using high speed digital inline holography (DIH), all the individual bubbles shed from the cavity are imaged downstream and are used to quantify the instantaneous gas leakage from the cavity. In general, the supercavity gas leakage exhibits significant fluctuations under all closure types with the instantaneous gas leakage rate spiking up to 20 times of the ventilation input under RJ and QV closures. The magnitude and occurrence rate of such excessive leakage is quantified using normalized excessive gas leakage $\hat{Q}_E$ (i.e., the relative percentage of instantaneous leakage above the ventilation input) and local Strouhal number $St_m$ (corresponding to the frequency of the occurrence of leakage maxima). As the supercavity transitions from RJ, to QV, TV, and PTV with increasing ventilation or decreasing tunnel speed, the $\hat{Q}_E$ decreases sharply from RJ to QV, plateaus from QV to TV, and drops again from TV to PTV. Correspondingly, the histogram of $St_m$ first exhibits a double peak distribution under RJ, migrates to a single peak mode under QV and TV, and eventually transitions to a distribution with a broadened peak at a higher frequency under PTV. These trends can be explained by the changing flow instabilities associated with gas leakage mechanisms under different closures. Specifically, the RJ cavity has two gas leakage modes, i.e., re-entrant jet impingement ($Q_{RJ}$) and vortex tube gas leakage ($Q_{VT}$), resulting in two characteristic frequencies in the $St_m$ distribution one coincident with the frequency of re-entrant jet impingement of the cavity gas-liquid interface and the other near the frequency of vortex tube breakup. The $Q_{VT}$ becomes the dominant gas leakage mode for QV and TV cavities causing $St_m$ to shift to a single peak distribution. The gas leakage mechanism for PTV cavity differs from the previous cases and is driven primarily by the bubble pocket shed-off ($Q_{BS}$) due to the pulsation of supercavity, which leads to a $St_m$ distribution with the peak matching the natural frequency of cavity pulsation. The change in tunnel speed and ventilation rate can shift the dominance of different gas leakage mechanisms, giving rise to the corresponding changes in $\hat{Q}_E$ and $St_m$. Particularly, we find that the supercavity under transitional closure types usually has high $\hat{Q}_E$, which can be attributed to the kicking in of additional instabilities associated with new gas leakage mechanisms. To characterize the controllability of a supercavity through ventilation adjustment, we introduce two metrics, i.e., normalized standard deviation of $Q_E$ to quantify the relative change of ventilation needed to compensate the change of extra gas loss and the approximate entropy of the simplified gas leakage signal to evaluate the predictability of the occurrence of excessive gas leakage. Based on the maps of these two metrics across different tunnel speed and ventilation conditions, we suggest that the supercavity operating under TV closure with moderate ventilation is optimal for ventilation-based controls.

Our study provides the first experimental investigation of supercavity gas leakage associated with inherent closure/cavity instability under different closure conditions. The excessive gas leakage spikes revealed in our study can cause significant disturbance to the internal gas flow and pressure distribution of a supercavity, yielding a detrimental impact on cavity stability. Nevertheless, our study suggests that the impact of such leakage spikes can be mitigated by operating the supercavity under specific closure and ventilation conditions and implementing appropriate ventilation controls. Specifically, both the rate and site of ventilation can be adjusted to compensate sudden increase of gas loss and minimize its impact on cavity stability. This type of ventilation adjustment can be achieved through a close loop control system using the information from the pressure sensors located on a supercavitating device. Our study provides some fundamental understanding and assessment of such ventilation-based controllability under different cavity closure and operational conditions.

In the end, we would like to caution the readers that our experiments were only conducted on a limited range of tunnel speeds and ventilation conditions, which is constrained by our facility and DIH measurement setup. However, the fundamental understanding gained from our work can be generalized to broader settings. Moreover, the benchmark data of supercavity bubbly wake and gas leakage generated from our study can be used for improving high-fidelity numerical models of ventilated supercavity and designing full-scale experiments to evaluate supercavity gas leakage behaviors under more practical settings.

## Acknowledgements

This work is supported by the Office of Naval Research (Program Manager, Dr. Thomas Fu) under Grant No. N000141612755. We thank Erik Steen, Christopher Ellis and Jim Tucker for their mechanical supports and advice during the water tunnel experiments.